# Anisotropic Gauge Theories: A Numerical Study of the Fu–Nielsen Model

Arjan Hulsebos[a]

[a]FB08, University of Wuppertal, Gaußstraße 20, 42119 Wuppertal, BRD.

We study numerically 4+1 dimensional U(1) pure gauge theory.

## 1. Introduction

Most studies concerning gauge theories only involve *isotropic* gauge theories, and for good reasons. It would be very surprising indeed to widness quarks deconfine by merely turning your head. Nevertheless, in some instances it can be useful to study anisotropic gauge theories. For instance, consider anisotropic U(1) gauge theory in d+D dimensions. By increasing the coupling $g_D^2$, while keeping $g_d^2$ fixed, the coupling involving plaquettes having links in the D dimensional subspace, we can imagine that at a certain point the coupling becomes so strong that confinement occurs in this subspace. In this way, we would have found a model for dimensional reduction.

This idea was indeed the starting point for Fu and Nielsen to study anisotropic U(1) gauge theories. In [1], they studied 4+1 dimensional U(1) gauge theory in the mean field approach. There, they found a new, 'layered' phase. This phase is Coulombic in the 'regular' four dimensions, but is confining in the fifth dimension.

The reason to re-examine this model is the Kaplan model: Wilson fermions coupled to a domain wall in a (spurious) fifth dimension [2]. When incorporating gauge fields, we might try anisotropic gauge couplings in order to do away with this extra dimension. We studied this model in ref. [3]. Here we will present some results from a numerical study of this model, and comment on the order of the phase transitions.

## 2. The model and the method

The action of 4+1 dimensional anisotropic U(1) gauge theory is given by

$$S_G = \beta \sum_{1\leq\mu<\nu\leq 4,x} (1-\Re U_{\mu\nu}(x)) + \beta' \sum_{x,\,\mu\leq 4} (1-\Re U_{\mu 5}(x)) \qquad (1)$$

This model was studied in mean field in [1,3,4]. The results are as follows. Three phases were identified: a Coulombic phase (C), a confining phase (CF) and a layered phase (L). The phase diagram is presented in figure 1. Mean field predicts 1st order phase transitions for the CF-L and the CF-C phase transition. The L-C phase transition is predicted to be 2nd order [4].

We will study this model numerically by means of a 5-hit Metropolis algorithm with an adaptable, site dependent stepsize. The lattice size for the results presented here was $8^5$. We performed several thermal runs through the various phase

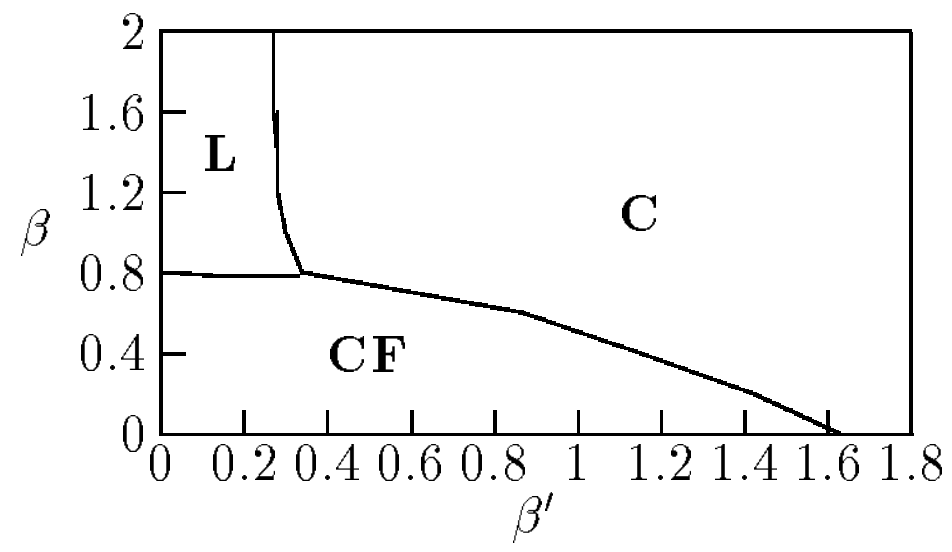

b

Figure 1. Phase Diagram of the five dimensional anisotropic $U(1)$ pure gauge theory using mean field techniques.

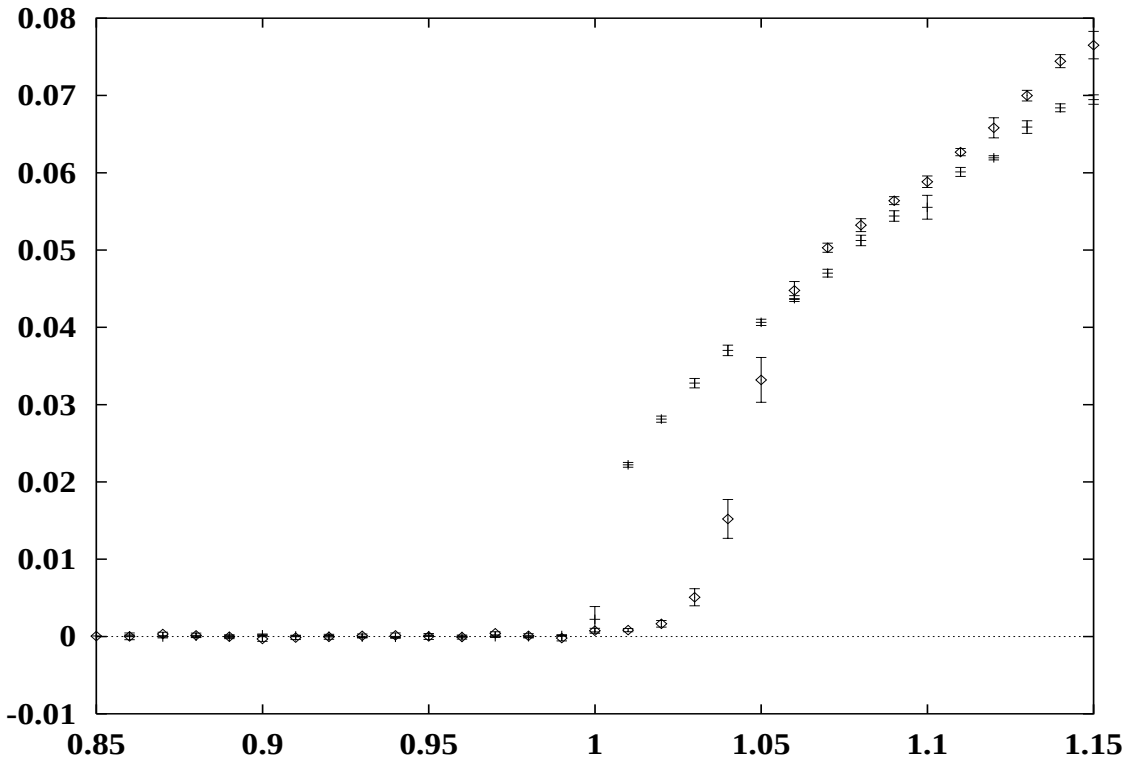

Figure 2. Polyakov line correlator (2) vs. $\beta$ at $\beta' = 0.20$ (CF–L).

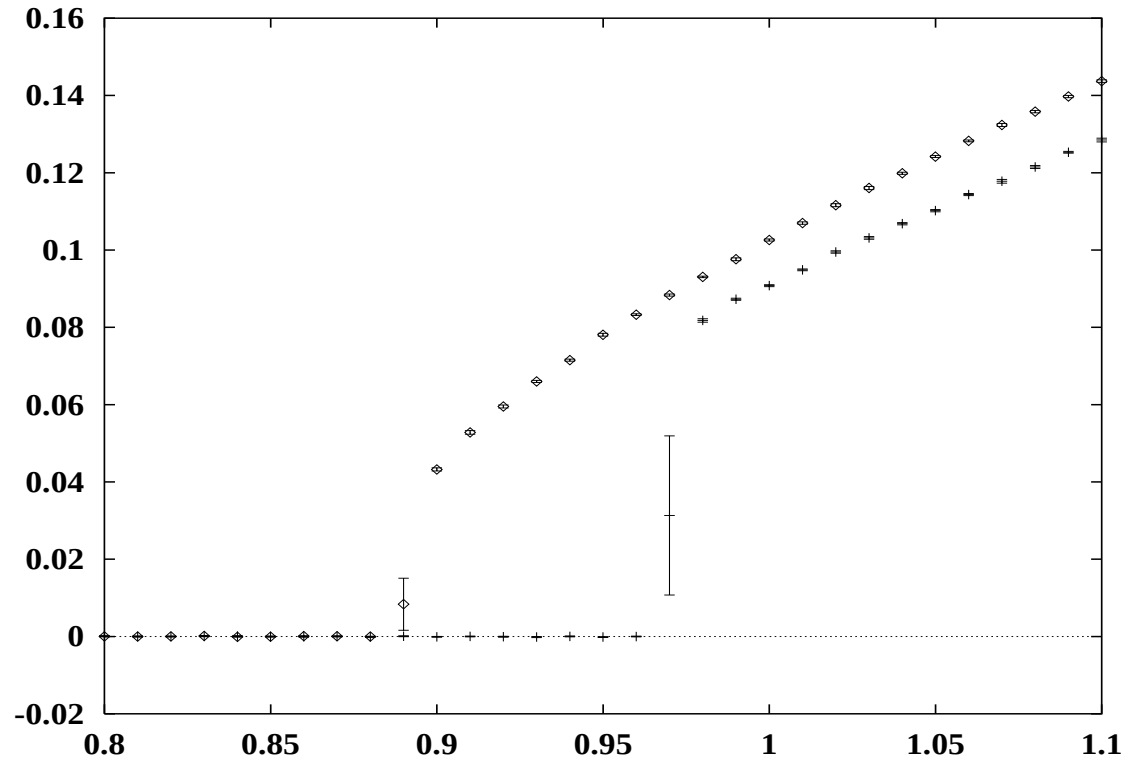

Figure 3. Polyakov line correlator (2) vs. $\beta$ at $\beta' = 0.50$ (CF–C).

transitions. A run consisted of 5k updates at the starting $(\beta, \beta')$ pair, followed by 1k updates and measurements. After these 1k updates, either $\beta$ or $\beta'$ was increased by 0.01, and another 1k measurements and updates were made without rethermalization. After reaching the final $(\beta, \beta')$ pair, the process was reversed until we arrived at the starting $(\beta, \beta')$ pair. Among other observables, we measured the following Polyakov loop correlators:

$$P_{\rm sp} = \left\langle \frac{1}{12V} \sum_{\substack{x,\mu,\nu \\ \mu,\nu<5 \\ \mu\neq\nu}} \Re\Big(p_\mu^\dagger(x) p_\mu(x + \tfrac{N}{2}\hat{\nu})\Big) \right\rangle \quad (2)$$

$$P_5 = \left\langle \frac{1}{4V} \sum_{\substack{x,\mu \\ \mu<5}} \Re\Big(p_5^\dagger(x) p_5(x + \tfrac{N}{2}\hat{\mu})\Big) \right\rangle \quad (3)$$

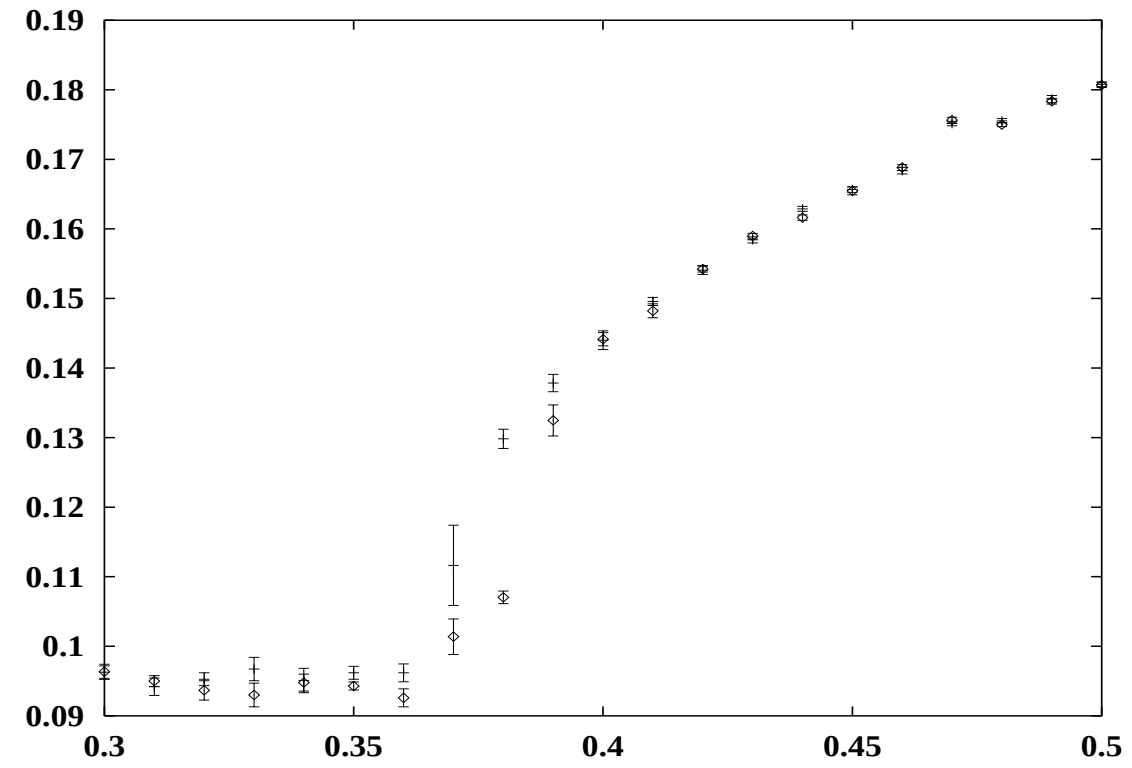

vspace*-12pt

Figure 4. Polyakov line correlator (2) vs. $\beta'$ at $\beta = 1.20$ (L–C).

## 3. Results

The results for (2) are displayed in figures 2 – 4.

The discrepancy between the two sets of data in figure 3 for $\beta > 0.98$ is due to the fact that $< p_\mu \neq 0$ in this region, due to (very) large autocorrelations in this observable. After correcting for this fact, we find agreement up to deviations of 2–3 $\sigma$.

From these pictures, it will be clear that the CF–C phase transition is 1st order. The L–C phase transition turns out to be 2nd order. The CF–L phase transition is a little more troublesome. In order to study this phase transition, we first devised a variant of the overrelaxation algorithm.

By noting that in general overrelaxation leaves the *action* invariant, we take the following update to be an overrelaxed update of link $U_\mu(x)$:

$$U_\mu(x) \to U'_\mu(x) = F_\mu^\dagger(x) U_\mu^\dagger(x) F_\mu^\dagger(x) \quad (4)$$

where $F_\mu(x) = f_\mu(x)/||f_\mu(x)||$, and

$$f_\mu(x) = \sum_{\nu\neq\mu} \beta_{\mu\nu}[U_\nu(x)U_\mu(x+\hat{\nu})U_\nu^\dagger(x+\hat{\mu}) + U_\nu^\dagger(x-\hat{\nu})U_\mu(x-\hat{\nu})U_\nu(x-\hat{\nu}+\hat{\mu})].$$

Using this version of overrelaxation, we were able to reduce autocorrelations considerably. We are now able to study more carefully the order of the CF–L phase transition. This was done in the following way. We made two runs at $\beta' = 0.20, \beta = 1.02$. The first run had an ordered starting configuration ($U_\mu = 1$ throughout the lattice). The second one had a random configuration as its stating point. We measured the above mentioned Polyakov line correlators, as well as Wilson loops up to $7\times7$ after 5 updates. One update consisted in this case of a 5-hit Metropolis sweep, followed by 5 overrelaxed sweeps for the $U_5$ links. For all observables, we noticed that after 40 measurements, that is after 200 updates, the two signals became indistiguishable, i.e. we could not find a two state signal. Therefore, we may conclude that it is unlikely that the CF–L phase transition is of 1st order, and it is very likely of *2nd* order. This should be contrasted with the CF–C phase transition. There, starting from cold and hot configurations, the two signals remained seperated for at least 50k updates, even on $4^5$ lattices.

## 4. Conclusions

We have studied 4+1 dimensional anisotropic U(1) gauge theory. We found that there are strong indications that the CF–L phase transition is of 2nd order, in contrast to the mean field results. In the process, we have extended the overrelaxed algorithm to deal with anisotropic systems.

## Acknowledgements

We would like to thank Chris Korthals-Altes and Stam Nicolis for useful discussions. This work was supported by EC contract SC1 *CT91-0642.